# Tunable Magnetless Optical Isolation with Twisted Weyl Semimetals

Vladislav Chistyakov[1], Viktar S. Asadchy[2], Shanhui Fan[3], Andrea Alù[4], and Alex Krasnok[5,*]

[1]*Saint-Petersburg, 191002, Russia*

[2]*Department of Electronics and Nanoengineering, Aalto University, 02150, Espoo, Finland*

[3]*Ginzton Laboratory and Department of Electrical Engineering, Stanford University, Stanford, CA, 94305, USA*

[4]*Photonics Initiative, Advanced Science Research Center, City University of New York, New York, NY, USA*

[5]*Department of Electrical and Computer Engineering, Florida International University, Miami, FL 33174, USA*

[*]*e-mail:* akrasnok@fiu.edu

**Abstract**

Weyl semimetals hold great promise in revolutionizing nonreciprocal optical components due to their unique topological properties. By exhibiting nonreciprocal magneto-optical effects without necessitating an external magnetic field, these materials offer remarkable miniaturization opportunities and reduced energy consumption. However, their intrinsic topological robustness poses a challenge for applications demanding tunability. In this work, we introduce an innovative approach to enhance the tunability of their response, utilizing multilayered configurations of twisted anisotropic Weyl semimetals. Our design enables controlled and reversible isolation by adjusting the twist angle between the anisotropic layers. When implemented in the Faraday geometry within the mid-IR frequency range, our design delivers impressive isolation, exceeding 50 dB, while maintaining a minimal insertion loss of just 0.33 dB. Moreover, the in-plane anisotropy of Weyl semimetals eliminates one or both polarizers of a conventional isolator geometry, significantly reducing the overall dimensions. These results set the stage for creating highly adaptable, ultra-compact optical isolators that can propel the fields of integrated photonics and quantum technology applications to new heights.

***Introduction.*** – Weyl semimetals (WSs) represent a fascinating 3D topological phase of matter, characterized by low-energy excitations that follow the Weyl equation. These materials contain even numbers of Weyl nodes within their energy band structure, which carry quantized topological charges [1–12]. Quasiparticles near these nodes resemble Weyl fermions in high-energy physics, displaying linear dispersion and distinct chirality. Acting as monopoles of Berry curvature, Weyl nodes remain stable under perturbations. WSs arise from Dirac cone states when either inversion (P) or time-reversal (T) symmetry is broken, and their resilience arises since a Weyl node can not be eliminated unless through annihilation with another Weyl node with opposite chirality [13]. Both electronic WSs and their photonic analogues have been experimentally verified, demonstrating unique optical properties such as the anomalous Hall effect, chiral magnetic effect, and giant magneto-optical Faraday/Kerr effects [9,10,14–21]. These properties give rise to various nonreciprocal phenomena, including nonreciprocal wave transfer, surface plasmons, and thermal emitters, making WSs valuable for both fundamental research and practical applications [22–26].

Nonreciprocal light transfer is crucial in photonics [27–29] and quantum technologies [30,31]. In photonics, isolators are indispensable for creating unidirectional optical circuits, reducing reflections, and eliminating multipath interference in communication channels [32,33]. Similarly, in quantum technology, isolators play a vital role in safeguarding quantum circuits from noise during readout. Nonreciprocal components primarily rely on the magneto-optical effect found in ferrite materials [32,34,35]. Unfortunately, these components are bulky, minimally tunable, and incompatible with planar technologies such as integrated photonics and transmission-line quantum circuits. Alternative approaches, including the use of two-dimensional magnetic materials [36] and time modulation [37], have been explored extensively in recent years. However, they still encounter tunability, bandwidth, and energy consumption limitations. Recent studies have shown that large optical isolation can be achieved by designing photonic structures based on Weyl semimetals [22,26,38,39]. The strong nonreciprocity in these materials arises from the anomalous Hall effect caused by Weyl node separation [26,40,41]. This mechanism differs fundamentally from the cyclotron mechanism in magneto-optical materials, and it does not necessitate an external magnetic field or static magnetization. Nonetheless, the inherent topological properties of these materials present challenges for practical applications that demand tunability.

In this work, we propose a novel approach to tunable optical isolation that utilizes multilayered structures with twisted optically anisotropic bias-free Weyl semimetals. The in-plane anisotropy can be either intrinsic [42,43] or artificially introduced by surface corrugation on the WS surface. Our approach enables highly efficient tuning of both direction and value of isolation by relative rotation of WSs layers. Moreover, in-plane anisotropy eliminates the need for one or both polarizers, significantly simplifying and downsizing the entire structure. Implemented within the Faraday geometry in the mid-IR frequency range, our design comprises two anisotropic WS layers separated by a dielectric medium. This configuration achieves exceptional isolation performance, surpassing 50 dB while maintaining an ultra-low insertion loss of just 0.33 dB.

*Results and discussion.* – **Fig. 1(a)** demonstrates a design that leverages the Faraday geometry. This setup incorporates a three-layer isolator structure composed of two layers of WSs separated by a dielectric with $\varepsilon_{\text{diel}} = 5$, and two linear polarizers angled at $45°$. The dielectric layer supports a Fabry-Perot resonance, with electric field mainly concentrated in this layer. This strategic arrangement effectively suppresses the losses incurred in WSs, thus increasing the overall efficiency of the setup. In the spectral range of interest, we have a wide variety of virtually lossless materials of such permittivity, including chalcogenide glasses (e.g., $As_2Se_3$, $As_2S_3$), titanium dioxide ($TiO_2$), silicon nitride ($Si_3N_4$).

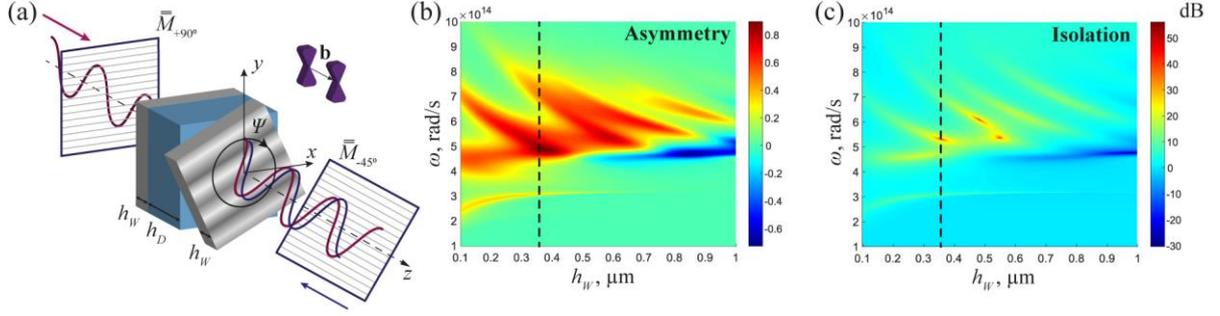

**Figure 1.** (a) Geometry of a Faraday isolator based on a dielectric ($\varepsilon_{\text{diel}} = 5$) sandwiched between two anisotropic WS slabs. Linear polarizers are represented by silver grids, while wavy curves depict the electric fields of linearly polarized waves moving in the $+z$ and $-z$ directions. The $+z$ direction is along the separation of Weyl nodes in momentum space **b**. Matrices $\overline{\overline{M}}_{+90} = 90°$ and $\overline{\overline{M}}_{-45} = -45°$ stand for the Jones matrices of the two polarizers. Rotation tuning is achieved by rotating the rightmost WS layer by angle $\Psi$. (b), (c) Asymmetry $\Delta T = |T_{+z}|^2 - |T_{-z}|^2$ and isolation $I = 10\lg(|T_{+z}|^2 / |T_{-z}|^2)$ as a function of frequency and thickness of the WS for $\Psi = 0°$. $|T_{+z}|^2$ ($|T_{-z}|^2$) is the sum of transmission of all polarization components, including the cross-polarisation, $|T_{+z}|^2 = \sum_{\substack{i=x,y \\ j=x,y}} |T_{+z}^{ij}|^2$ in the $+z$ ($-z$) direction.

Polarizers are essential components in optical isolators. The input polarizer ensures the incoming light is linearly polarized in a specific orientation. The Faraday rotator, in our case, the three-layer structure, rotates the plane of polarization by 45 degrees when the light passes through it. The output polarizer is oriented to allow light with the rotated polarization to pass through. When light attempts to travel in the reverse direction, it first passes through the output polarizer, which does not change its polarization. However, when it reaches the Faraday rotator, the polarization is rotated by another 45 degrees. This results in a total rotation of 90 degrees compared to the original forward-traveling light. Since the input polarizer is oriented to block this rotated polarization, the light cannot pass through, effectively achieving the one-way transmission of light.

Upon a comprehensive examination of existing theoretical and experimental research on the optical characteristics of WSs, we choose to utilize the relative permittivity tensor [24,44]

$$\varepsilon_{WS} = \begin{pmatrix} \varepsilon_d & i\varepsilon_a & 0 \\ -i\varepsilon_a & \varepsilon_d' & 0 \\ 0 & 0 & \varepsilon_d \end{pmatrix} \quad (1)$$

and unity magnetic permeability. The Weyl nodes splitting in momentum space by the vector **b** gives rise to the off-diagonal components $\varepsilon_a = be^2 / 2\pi^2 \hbar \omega$ enabling the nonreciprocal behavior. As a result, WSs support magneto-optical effects when light propagates along this vector ($\mathbf{k} \parallel \mathbf{b}$), where **k** denotes the light wave vector. It is noteworthy that, although we illustrate our focus on the case of two Weyl nodes as seen in EuCd$_2$As$_2$ [34], our consideration can be generalized to other types of WSs. To determine the diagonal components, we employ the Kubo-Greenwood formalism [44,45]. Remarkably, $\varepsilon_a$ and $\varepsilon_d$ exhibit values on the order of unity across the entire spectrum, indicating an exceptionally large magneto-optical parameter $\varepsilon_a/\varepsilon_d \approx 1$. This result represents a significant improvement over conventional magneto-optical materials in IR [26,27]. Further details can be found in the supplementary material (SM). Our model utilizes parameters close to experimental values, as presented in ref. [24] and SM.

We introduce the anisotropy in the diagonal terms of the tensor by $\varepsilon'_d \neq \varepsilon_d$. **Fig. S1** in SM illustrates the frequency dispersion of the real part of the $\varepsilon_d$, $\varepsilon_a$, and $\varepsilon'_d$. Anisotropy is invoked by shifting the plasma frequency for the y-component of the dielectric tensor [46,47]. For example, the in-plane optical anisotropy can be intrinsic in noncentrosymmetric WS [42,43]. Another way to in-plane anisotropy involves creating corrugations with subwavelength granularity to achieve an anisotropic effective permittivity. This approach has been recently implemented in developing hyperbolic metasurfaces [46]. We describe this approach and its results in **Fig. 4**. The in-plane anisotropy of the material presents a compelling opportunity to exert control over the optical response by manipulating the relative orientation of its layers. This concept has been recently explored in anisotropic metasurfaces and natural materials [48–52]. In our structures, the rotation tuning is achieved by rotating the rightmost WS layer by angle $\Psi$.

The calculation of transmission coefficients is carried out using the generalized T-matrix formalism [53–55] and reinforced by rigorous full-wave numerical simulations. Jones matrices are utilized to account for the linear polarizers. The thickness of the dielectric $h_{\text{diel}} = 0.85$ μm enables the Fabry-Perot mode at the plasma frequency of WS ($\text{Re}(\varepsilon_d) = 0$), $\Omega_p = 4.2 \times 10^{14}$ rad/s. The thickness of the WSs $h_W = 0.37$ μm was optimized to ensure maximum asymmetry $\Delta T = |T_{+z}|^2 - |T_{-z}|^2$ and isolation $I = 10\lg\left(|T_{+z}|^2/|T_{-z}|^2\right)$, **Figs. 1(b),(c)**. Here, $|T_{+z}|^2$ ($|T_{-z}|^2$) is the sum of the transmission of all polarization components, including the cross-polarisation, $|T_{+z}|^2 = \sum_{\substack{i=x,y \\ j=x,y}} |T_{+z}^{ij}|^2$ in the $+z$ ($-z$) direction.

The maximum isolation for selected optimized parameters reaches 54 dB at the frequency $\omega = 5.4 \times 10^{14}$ rad/s ($\lambda = 3.5$ μm), **Fig. 1(c)**. The total length of the structure without polarizers is $L = 1.59$ μm, which is only $0.45\lambda$ at the maximum isolation.

**Figs. 2(a),(b)** show the forward (+z) and backward (-z) transmittance of the structure as a function of the rotation angle between the layers and frequency. The system exhibits excellent isolation properties with complete transmission in the forward direction and vanishing transmission in the backward direction. The results of the isolation spectrum are shown in **Fig. 2(c)**. The isolation is tuned by adjusting the rotation angle, reaching a maximum of 60dB at $\Psi = 38°$ and $\omega = 4.9 \times 10^{14}$ rad/s (

$\lambda = 3.8$ μm), **Fig. 2(d)**. This excellent isolation occurs for a remarkably thin isolator of length $L = 0.45\lambda$. The relative rotation of the WS layers changes the amplitude and frequency of the isolation, providing a unique approach to tunable optical isolators, **Figs. 2(c),(d)**.

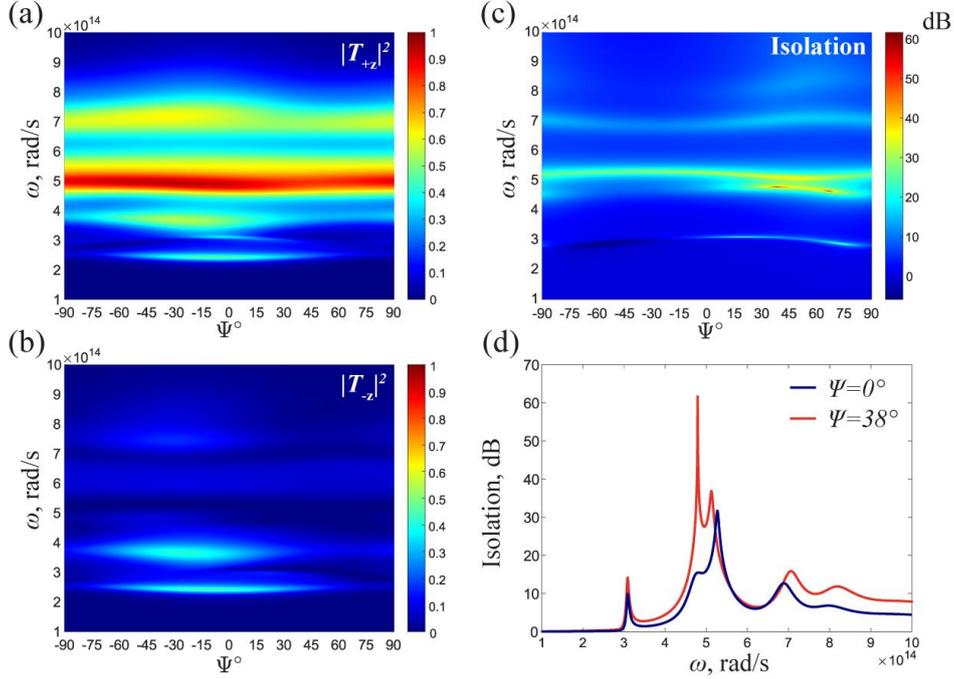

**Figure 2.** (a),(b) Transmittance as a function of frequency $\omega$ and rotation angle $\Psi$ in $+z$ and $-z$ directions, respectively. (c) Isolation versus frequency and rotation angle. (d) Isolation coefficient $I = 10\lg\left(|T_{+z}|^2 / |T_{-z}|^2\right)$ as a function of frequency at the twist angle $0°$ and $38°$.

Now we demonstrate that the anisotropy of WSs in this geometry can operate as a polarizer, eliminating the $-45°$ polarizer and making the design even more compact. **Fig. 3(a)** shows the geometry of the proposed compact isolator with only one input polarization beam splitter. The thickness of the first layer of the WS is adjusted to $h_w^1 = 0.92$ μm to reach maximum isolation of 20 dB at $\omega = 3 \times 10^{14}$ rad/s (the plasma frequency for $\varepsilon_d'$) and at the twisting angle $\Psi = 50°$, **Fig. 3(b)**. While the lack of a polarizer requires the first layer of the WS to be thicker to achieve high isolation, the design can be thinner in practice because polarizers are typically positioned several wavelengths away from the structure. Other geometrical parameters are the same as in **Figs. 1,2**. **Fig. 3(c)** shows the isolation spectrum as a function of the relative rotation and frequency. At the plasma frequency $\omega = 3 \times 10^{14}$ rad/s of the anisotropic component $\varepsilon_d'$, an isolation resonance arises, which is tuned with rotation, reaching an extreme at the twisting angle $\Psi = \pm 50°$ where the isolation reaches $\pm 20$ dB, **Fig. 3(d)**. Therefore, at this rotation angle, the maximum effect is achieved when the polarization of the backward wave turns into a horizontal one and is blocked by the incoming polarizer. This design allows for both spectral control of isolation and the ability to control its direction, including the reversal of isolation.

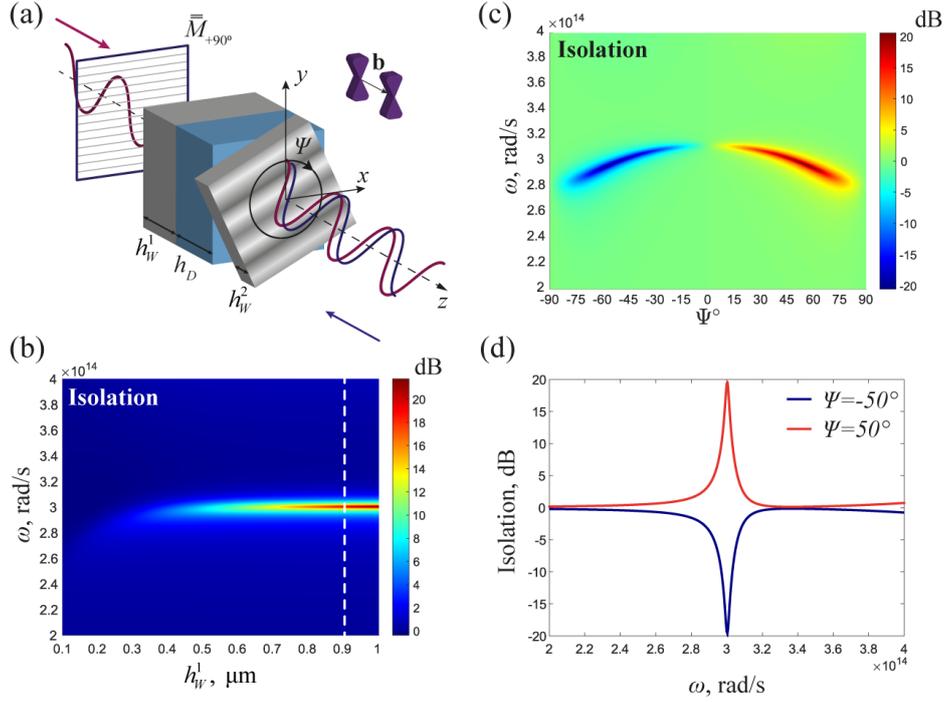

**Figure 3.** (a) Isolator geometry with one polarizer. (b) Isolation spectrum $I = 10\lg\left(|T_{+z}|^2 / |T_{-z}|^2\right)$ as a function of frequency and thickness of the Weyl semimetal at a rotation angle $\Psi = 50°$. (e) Isolation as a function of frequency and angle of rotation. $|T_{+z}|^2$ ($|T_{-z}|^2$) is the sum of transmission of all polarization components, including the cross-polarisation, $|T_{+z}|^2 = \sum_{\substack{i=x,y \\ j=x,y}} |T_{+z}^{ij}|^2$ in the $+z$ ($-z$) direction.

Next, we examine how the anisotropy of WSs can be attained by incorporating periodic corrugations of the WSs' surface, as depicted in **Fig. 4(a)**. Here the grooves in the corrugated slabs are solely responsible for the anisotropy. We validate our analytical results through full-wave simulations using CST Microwave Studio, where unit-cell boundary conditions and Floquet's ports simulate the periodic structure. To achieve a metasurface regime, we select a corrugation width $a$ at least ten times smaller than the wavelength at the resonant plasma frequency $\Omega_p = 4.2 \times 10^{14}$ rad/s. We choose $a = 0.11$ µm, with a depth of $t = 0.47$ µm and $p = 0.14$ µm. To achieve higher isolation, we adjust the thickness of the WS in simulations to $h_W = 0.7$ µm resulting in a total isolator thickness of $L = 2.34$ µm. **Fig. 4(c)** depicts the $|E_x|^2$ distribution in the yz-plane for the structure with a relative twist angle of $\Psi = 0°$ and frequency $\omega = 2.7 \times 10^{14}$ rad/s. As expected from our analytical analysis, this scenario is reciprocal and symmetric, $\Delta T = |T_{+z}^{xx}|^2 - |T_{-z}^{xx}|^2 = 0$. In **Fig. 4(d)**, we present the $|E_x|^2$ field distribution at a relative twist angle of $\Psi = 90°$. The figure reveals that the structure transmits x polarisation of radiation ($|T_{+z}^{xx}|^2$) in $+z$, while almost entirely suppressing it in the opposite direction ($|T_{-z}^{xx}|^2 \approx 0$). The asymmetry coefficient obtained reaches 85% at $\omega = 2.7 \times 10^{14}$ rad/s ($L = 3.9$ µm), in good agreement with the analytical results, **Fig. 4(b)**.

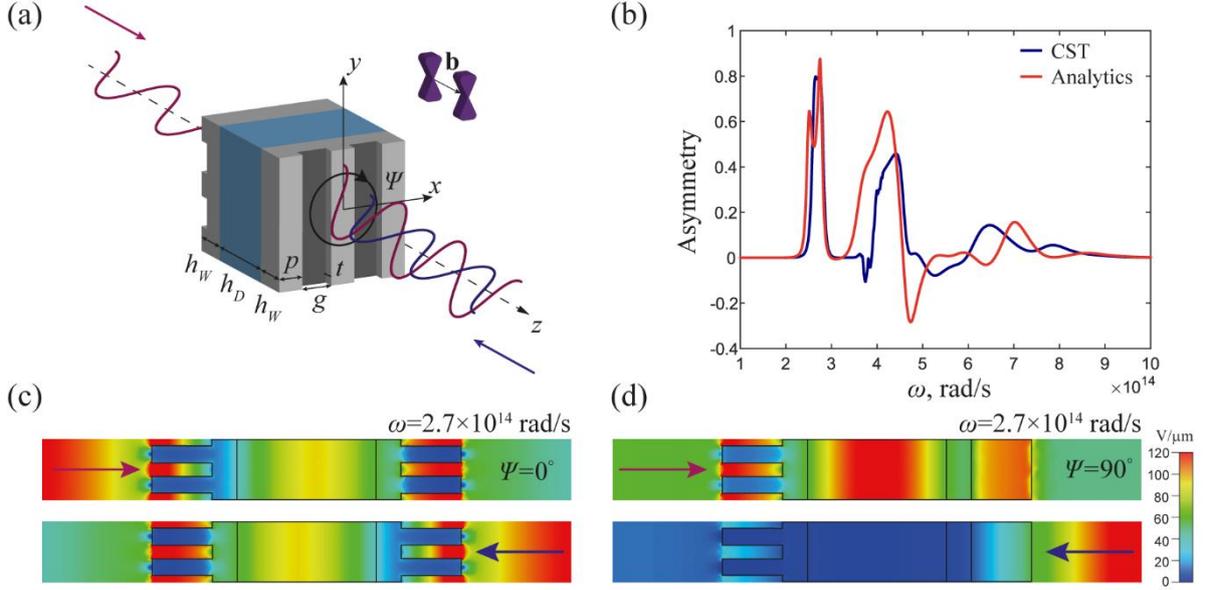

**Figure 4.** (a) Faraday rotator geometry with corrugated metasurface surface and without polarizers. (b) Transmittance asymmetry $\Delta T = |T^{xx}_{+z}|^2 - |T^{xx}_{-z}|^2$ as a function of frequency at the twist angle $\Psi = 90°$. The analytical results are obtained using the T-matrix approach with the effective material parameters for the corrugated metasurfaces extracted from full-wave simulations in CST Microwave Studio. (c),(d) Absolute value of $|E_x|^2$ distribution of the isolator in the yz-plane at the twist angle (a) $\Psi = 0°$ and (c) $\Psi = 90°$ and frequency $\omega = 2.7 \times 10^{14}$ rad/s, for the forward (upper) and backward (lower) propagation.

The current design is a Faraday rotator that works as a polarized isolator, exhibiting significant isolation solely for specific polarizations of forward and backward waves. Without polarizers enabling polarization selection, waves with arbitrary incoming polarization cannot be blocked entirely. Nonetheless, this limitation can be overcome by increasing the thicknesses of the anisotropic WS layers or using at least one polarizer.

***Conclusion.*** -- This work presents a novel approach to tunable optical isolation based on twisted bilayered Weyl semimetals. We have demonstrated that our approach has enabled highly efficient tuning of both direction and value of isolation by utilizing the relative rotation of bilayered Weyl semimetals. The structure of the Faraday isolator with two polarizers consists of two anisotropic Weyl semimetals layers separated by a dielectric. We have shown that this approach has achieved isolation exceeding 50 dB and an insertion loss as small as 0.33 dB. Moreover, we have found that the anisotropy of the Weyl semimetals has made it possible to eliminate one of the polarizers in the isolator, simplifying the model and making it more compact. The tunable properties, high isolation coefficient, and the absence of a requirement for an external magnetic field and ultrasmall sizes make our proposed design superior to other approaches based on magneto-optical or temporal modulation effects. We anticipate that further development of Weyl semimetals and reducing their material losses will lead to even better performance characteristics of these optical isolators.